\documentstyle[12pt,epsf]{article}

\fussy
\begin{document}
\newcommand{\p}{\partial}
\newcommand{\ls}{\left(}
\newcommand{\rs}{\right)}
\newcommand{\beq}{\begin{equation}}
\newcommand{\eeq}{\end{equation}}
\newcommand{\beqa}{\begin{eqnarray}}
\newcommand{\eeqa}{\end{eqnarray}}
\newcommand{\bdm}{\begin{displaymath}}
\newcommand{\edm}{\end{displaymath}}
\newcommand{\fps}{f_{\pi}^2 }
\newcommand{\mks}{m_{{\mathrm K}}^2 }
\newcommand{\ms}{m_{{\mathrm K}}^{*} }
\newcommand{\msq}{m_{{\mathrm K}}^{*2} }
\newcommand{\rhos}{\rho_{\mathrm s} }
\newcommand{\rhob}{\rho_{\mathrm B} }
\title{Chiral kaon dynamics in heavy ion collisions} 
\author{C. Fuchs, Amand Faessler, Z.S. Wang and T. Gross-Boelting\\
Institut f\"ur Theoretische Physik der 
Universit\"at T\"ubingen, \\
D-72076 T\"ubingen, Germany }
\date{}
\maketitle
\begin{abstract}
The influence of the chiral mean field on the collective motion of 
kaons in relativistic heavy ion reactions at SIS energies is 
investigated. We consider three types of collective motion, i.e. 
the transverse flow, the out-of-plane flow (squeeze-out) and the radial 
flow. The kaon dynamics is thereby described with a relativistic mean field 
as it originates form chiral lagrangiens. For the $K$ mesons 
inside the nuclear 
medium we adopt a covariant quasi-particle picture including scalar 
and vector fields and compare this to a treatment with a static potential 
like force. The comparison to the available data ($K^+$) measured by 
FOPI and KaoS strongly favor the existence of an in-medium potential. 
However, using full covariant dynamics makes it more difficult to 
describe the data which might indicate that the mean field level is not 
sufficient for a reliable description of the kaon dynamics. 
\end{abstract}
\section{Introduction}
In recent years strong efforts have been made towards a better understanding 
of the medium properties of kaons in dense hadronic matter. This feature 
is of particular relevance since the kaon mean field is related to 
chiral symmetry breaking \cite{kaplan86}. 
The in-medium effects give rise to an 
attractive scalar potential 
inside the nuclear medium which is in first order, i.e. in mean field 
approximation, proportional to the kaon-nucleon Sigma term 
$\Sigma_{\mathrm{KN}}$. A second part of 
the mean field originates from the interaction with vector mesons 
\cite{kaplan86,brown96,schaffner97}. The vector potential is 
repulsive for kaons $K^+$ and, due to G-parity conservation, attractive 
for antikaons $K^-$. A strong attractive potential for antikaons may 
also favor $K^-$ condensation at high nuclear densities and thus modifies 
the properties of neutron stars \cite{li97}. 

One has extensively searched for signatures of 
these kaon-nucleus potentials in heavy ion reactions at 
intermediate energies \cite{fopi95,kaos97,kaos98}. 
In particular the collective motion of kaons in the dense hadronic 
environment is expected to be influenced by such medium effects 
\cite{li97,ko95,cassing97,wang97,ko97a,wang98}. In this work we discuss 
the in-medium kaon dynamics for $K^+$ and $K^-$ with respect to the 
three prominent types of collective motion in heavy ion reactions, 
namely the radial flow \cite{wang98}, the emission out of the reaction 
plane (squeeze-out) \cite{wang98b} and the in-plane flow (transverse flow) 
\cite{ko95,wang97,fuchs98}. The interaction of the kaons with the dense 
hadronic medium is thereby described on the mean field level based 
on chiral models \cite{kaplan86,ko95}. Following Ref. 
\cite{fuchs98} we discuss effects which originate from a fully covariant 
description of the kaonic mean field in a relativistic quasi-particle 
picture and compare this approach to the standard treatment with 
static potentials.
\section{Covariant kaon dynamics}
Due to its relativistic origin, the kaon mean field has a typical 
relativistic scalar--vector type structure. For the nucleons such 
a structure is well known from Quantum Hadron Dynamics \cite{serot88}.  
This decomposition of the mean field 
is most naturally expressed by an absorption of the scalar and vector 
parts into effective masses and momenta, respectively, leading to 
a formalism of quasi-free particles inside the nuclear medium
\cite{serot88}. 

From the chiral Lagrangian \cite{kaplan86} the field equations for 
the $K^\pm$--mesons are derived from the 
Euler-Lagrange equations \cite{ko95} 
\beq
\left[ \partial_\mu \partial^\mu \pm \frac{3i}{4\fps} j_\mu \partial^\mu 
+ \left( \mks - \frac{\Sigma_{\mathrm{KN}}}{\fps} \rhos \right) 
\right] \phi_{\mathrm{K^\pm}} (x) = 0
\quad .
\label{kg1}
\eeq
Here the mean field approximation has already been applied. In Eq. 
(\ref{kg1}) $ \rhos$ is the baryon scalar density and $j_\mu$ the baryon 
four-vector current. Introducing the kaonic vector potential 
\beq 
V_\mu = \frac{3}{8\fps} j_\mu
\label{vpot1}
\eeq
Eq. (\ref{kg1}) can be rewritten in the form \cite{fuchs98}
\beqa
&&\left[ \left( \partial_\mu + i V_\mu \right)^2  + \msq \right] 
\phi_{\mathrm{K^+}} (x) = 0 
\label{kg2a}\\
&&\left[ \left( \partial_\mu - i V_\mu \right)^2  + \msq \right] 
\phi_{\mathrm{K^-}} (x) = 0 
\quad . 
\label{kg2b}
\eeqa
Thus, the vector field is introduced by minimal coupling 
into the Klein-Gordon equation. The effective mass $\ms$ of 
the kaon is then given by \cite{schaffner97,fuchs98}
\beq
\ms = \sqrt{ \mks - \frac{\Sigma_{\mathrm{KN}}}{\fps} \rhos 
     + V_\mu V^\mu }
\quad . 
\label{mstar1}
\eeq
\begin{figure}[t]
\begin{center}
\leavevmode
\epsfxsize = 18cm
\epsffile[0 100 550 430 ]{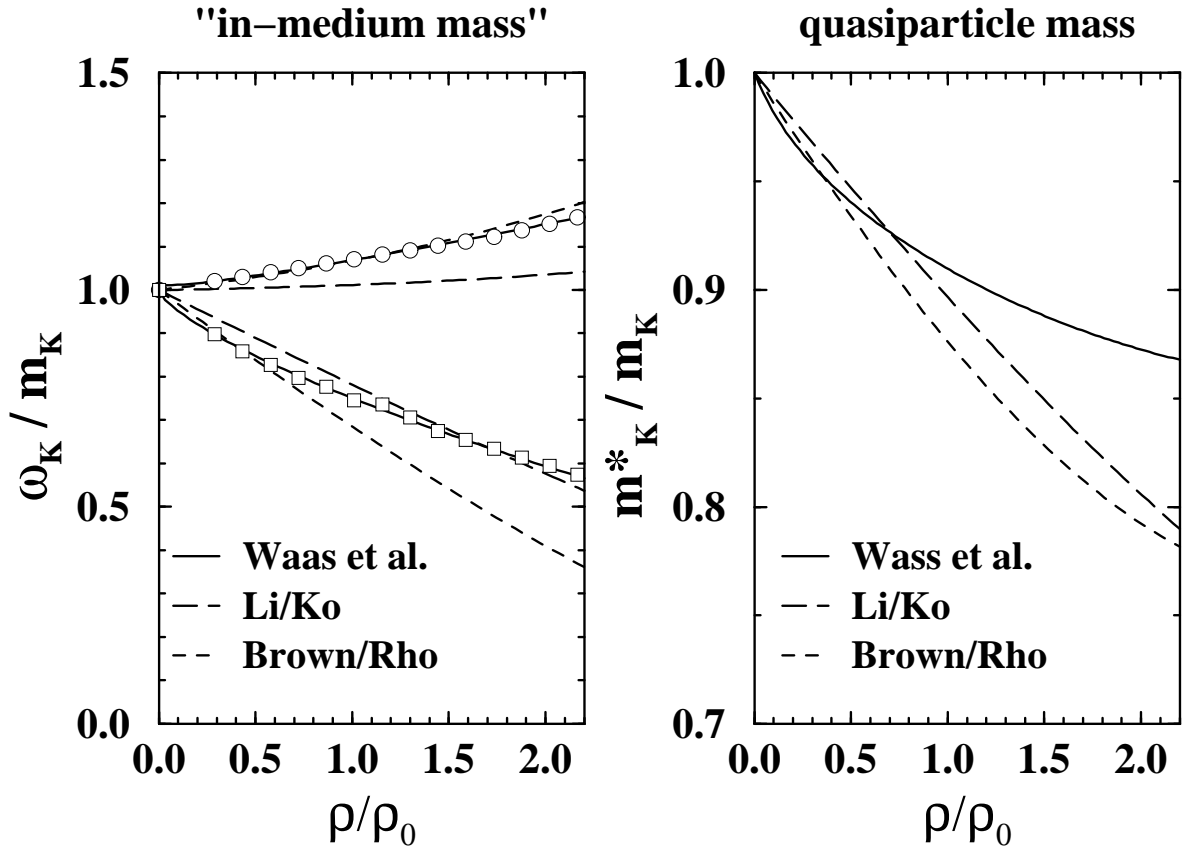}
\end{center}
\caption{{\it In-medium kaon mass in nuclear matter for various mean 
fields derived from chiral lagrangiens. The right panel shows 
the kaon energy at zero momentum which is often 
denoted as in-medium mass in the literature (the upper lines refer 
to $K^+$, the lower ones to $K^-$). The left panel shows the 
quasi-particle mass given by Eq. (\protect\ref{mstar1}) which is 
equal for $K^+$ and $K^-$.}  
}
\label{mstarfig}
\end{figure}
Due to the bosonic character, the coupling of the 
scalar field to the mass term is no longer linear 
as for the baryons but quadratic and contains an additional contribution 
originating from the vector field. The effective quasi-particle 
mass defined by Eq. (\ref{mstar1}) is a Lorentz scalar and 
is equal for $K^+$ and $K^-$. It should not be mixed up with 
the quantity, i.e. kaon energy at zero momentum 
$\omega ({\bf k}=0) = \ms \pm V_0$ for $K^\pm$, which is 
sometimes denoted as in-medium mass 
\cite{brown96,ko95,waas96,cassing97} and which determines the 
shift of the corresponding production thresholds. 
These two quantities, namely the energy at zero momentum 
and the in-medium quasiparticle mass $\ms$ are compared 
in Fig.1. For the modification of the kaon in-medium properties 
we parameterize results obtained form coupled channel 
calculation in chiral perturbation theory \cite{waas96} (ChPT). We  
also compare them to two more simple mean field models of the 
type of Eq. (\ref{kg1}) suggested by Li and Ko \cite{ko95} (MF) as 
well as by Brown and Rho \cite{brown96} (MF2). 

As can be seen from Fig. 1 the quasi-particle mass $\ms$ is 
equal for $K^+$ and $K^-$ and generally reduced inside the 
nuclear medium. However, in the approach of Ref. \cite{waas96} 
(Waas et al.) this reduction is much weaker than in the simple MF 
parameterization where the scalar field is in first order proportional 
to the scalar nucleon density.

Introducing an effective momentum $k^{*}_{\mu} = k_{\mu} \mp V_\mu $ 
for $K^+ (K^- )$, the Klein-Gordon 
equation (\ref{kg2a},\ref{kg2b}) reads in momentum space 
\beq
\left[ k^{*2} - \msq \right] \phi_{\mathrm{K}^\pm } (k) = 0
\label{kg3}
\eeq
which is just the mass-shell constraint for the quasi-particles 
inside the nuclear medium. These quasi-particles can now 
be treated like free particles. 
In nuclear matter at rest the spatial components of the 
vector potential vanish, i.e. ${\bf V} = 0$, and 
Eqs. (\ref{kg2a},\ref{kg2b}) reduce to the expression already 
given in Ref. \cite{ko95}. 

The covariant equations of motion for the kaons 
are obtained in the classical 
(testparticle) limit from the relativistic transport 
equation for the kaons which can be derived from Eqs. 
(\ref{kg2a},\ref{kg2b}). They are analogous to the 
corresponding relativistic equations for baryons and read  
\beqa
\frac{ d  q^\mu}{d\tau} = \frac{k^{*\mu}}{\ms}
\quad , \quad 
\frac{ d  k^{*\mu}}{d\tau} = \frac{k^{*}_{\nu}}{\ms} F^{\mu\nu} 
+\partial^\mu \ms
\quad . 
\label{como}
\eeqa
Here $q^\mu = (t,{\bf q})$ are the coordinates in Minkowski space 
and $F^{\mu\nu} = \partial^\mu  V^\nu - \partial^\nu  V^\mu $ is the 
field strength tensor for $K^+$. For $K^-$ where the vector field 
changes sign the equation of motion are identical, however, 
$F^{\mu\nu}$ has to be replaced by $-F^{\mu\nu}$. 
The structure of Eqs. (\ref{como}) may become 
more transparent considering only the spatial components
\beq
\frac{d {\bf k^*}}{d t} = - \frac{\ms}{E^*} 
\frac{\partial \ms }{\partial {\bf q}} \mp 
\frac{\partial V^0 }{\partial {\bf q}} 
\pm \frac{{\bf k}^*}{E^*} \times 
\left( \frac{\partial}{\partial {\bf q}} \times {\bf V} \right)
\label{lorentz}
\eeq
where the upper (lower) signs refer to $K^+$ ( $K^-$). 
The term proportional to the spatial component of the vector 
potential gives rise to a momentum dependence 
which can be attributed to a Lorentz force, i.e. 
the last term in Eq. (\ref{lorentz}). Such a velocity dependent 
$({\bf v} = {\bf k}^* / E^* )$ Lorentz force 
is a genuine feature of relativistic dynamics as soon 
as a vector field is involved. If the equations of motion are, 
however, derived from a static 
potential 
\beqa
 U({\bf k},\rho) = \omega({\bf k},\rho) - \omega_0 ({\bf k}) 
= \sqrt{{\bf k}^2 +  \mks - \frac{\Sigma_{\mathrm{KN}}}{\fps} \rhos 
     + V_{0}^2 } \pm V_0 - \sqrt{{\bf k}^2 +  \mks }
\label{pot}
\eeqa
as, e.g. in Refs. \cite{ko95,li97,cassing97}, the 
Lorentz-force like contribution is missing. The same holds for 
non-relativistic approaches \cite{wang97,wang98} where 
the Lorentz force has also not yet been taken into account. 
Such non-covariant treatments are formulated 
in terms of canonical momenta $k$ instead of kinetic 
momenta $k^*$ and then the equations of motion (\ref{lorentz}) 
read \cite{fuchs98,nantes98}
\beqa
\frac{d {\bf k}}{d t} = - \frac{\ms}{E^*} 
\frac{\partial \ms }{\partial {\bf q}} \mp 
\frac{\partial V^0 }{\partial {\bf q}} 
\pm {\bf v}_{i} \frac{\partial  {\bf V}_{i}}{\partial {\bf q}} 
= - \frac{\partial}{\partial {\bf q}} U({\bf k},\rho) 
\pm {\bf v}_{i} \frac{\partial  {\bf V}_{i}}{\partial {\bf q}} 
\label{lorentz2}
\eeqa
with ${\bf v} = {\bf k}^* / E^*$ the kaon velocity. Thus Eqs. 
(\ref{lorentz}) and (\ref{lorentz2}) are more general than the 
treatment in the static potential $U$, Eq.(\ref{pot}). 
Effects which arise from 
the full dynamics will be discussed in the following.  
\section{Collective flow of kaons}
\subsection{Radial flow}
One way to obtain information on the collective motion is to 
investigate particle multiplicities as a function of the transverse
mass $m_t$ \cite{wang98}. In the left panel of 
Fig. 2 the transverse mass spectrum of 
$K^+$ mesons emitted at midrapidity (-0.4$<$$y_{c.m.}$/$y_{proj}$$<$0.4) 
is shown for a central (b=3 fm) Au+Au reaction at 1 A.GeV. 
In the presence of a kaon potential (MF2, taken from \cite{brown96}) 
the kaon $m_t$ spectrum clearly exhibits a "shoulder-arm" shape 
which deviates from a pure thermal picture. On the other hand the 
calculation without any potential can be well described 
by a Boltzmann distribution, which is more or less a straight 
line if plotted logarithmically. Thus, the "shoulder-arm" 
structure is caused by the mean field 
rather than by a collective expansion of the kaon sources.
\begin{figure}[t]
\begin{center}
\leavevmode
\epsfxsize = 18cm
\epsffile[0 100 550 430 ]{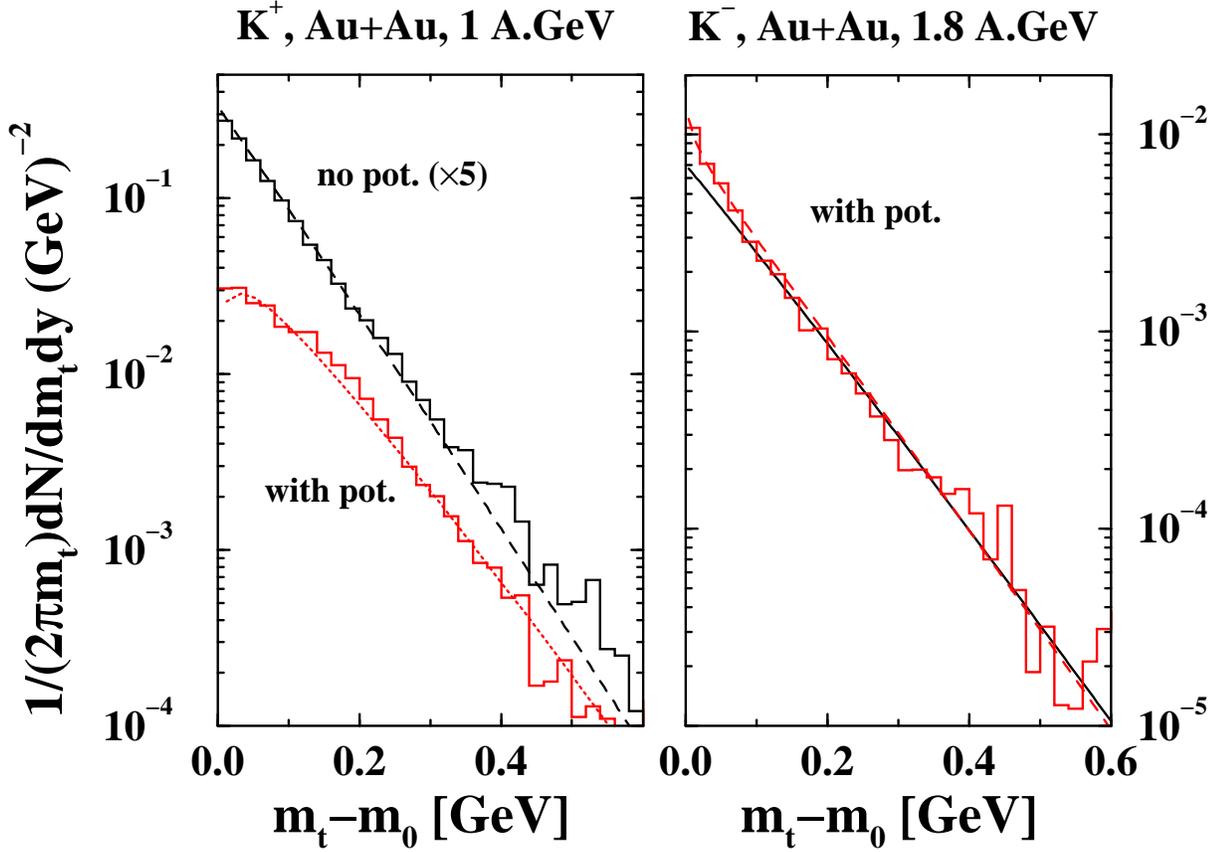}
\end{center}
\caption{{\it Influence of the in-medium potential on 
the kaon $m_t$ spectrum. The left panel shows two calculations for $K^+$ 
with and without repulsive potential for a central (b=3 fm) Au+Au reaction 
at 1 A.GeV. For a better representation the results without potential 
are multiplied by a factor of 5. The lines are Boltzmann fits to the 
calculations including (lower curve) and without (upper curve) 
a radial flow. The left panel shows the results 
for $K^-$ including the attractive 
potential in a semi-central (b=5 fm) 
Au+Au reaction at 1.8 A.GeV. The straight line is a Boltzmann fit 
without radial flow, the dashed line includes a virtual, inverse 
radial flow fit.} 
}
\label{radialfig}
\end{figure}
The kaons experience an acceleration due to the repulsive potential
as they propagate outwards from the participant region. We want to 
mention that these calculations have been performed with the QMD model 
where the kaon dynamics was treated non-covariantly, i.e. applying 
Eq. (\ref{lorentz2}) and neglecting thereby the contribution of 
the spatial vector field components. However, including these Lorentz 
forces we find that the results for the $m_t$ spectrum are hardly 
affected since the kaons stem from the equilibrated fireball zone 
where the collective motion of the nucleons relative to the kaons 
is rather small. To extract the collective component from the kaon spectrum,
we fit the QMD results including a common radial
velocity to the standard Boltzmann distribution
\beq
\frac{{d^3}N}{d{\phi}dy{m_t}d{m_t}} \sim e^{-(\frac{{\gamma}E}{T}-{\alpha})}
\{ {\gamma}^{2}E - {\gamma}{\alpha}T(\frac{E^2}{p^2}+1) + ({\alpha}T)^{2}\frac{
E^2}{p^2} \}\frac{\sqrt{({\gamma}E-{\alpha}T)^{2}-m^{2}}}{p}
\label{reswidth}
\eeq
where $E$ = $m_t$$coshy$, $p$ = $\sqrt{{p_t}^2+{m_t}^2sinh^2y}$,
$\alpha$ = $\gamma$$\beta$$p/T$, $\gamma$ = $(1-\beta^2)$$^{-1/2}$.
The fit yields $\beta=0.11$ and $T = 62 MeV$.

The attractive potential for $K^-$ leads on the other hand to 
an enhancement of low energetic particles and an concave spectrum 
seen on the right panel of Fig.2. Turning the sign of the flow 
velocity $\beta$ in Eq. (\ref{reswidth}) the calculation can 
also be described by an radial imploding source, i.e. by an 
virtual radial flow \cite{wang98}. Thus the effect of the kaon potential 
on the collective motion can clearly be distinguished from 
the thermal contribution.
\subsection{Squeeze-out}
In this subsection we study the azimuthal asymmetry of 
$K^+$ and $K^-$ emission \cite{wang98b}. At SIS energies shadowing dominates 
the reaction of nucleons and pions and leads to an enhanced particle  
emission out of the reaction plane. The phenomenon 
is usually called "squeeze-out". For $K^+$ mesons the shadowing effect 
can be expected to be small due to the moderate absorption cross 
section $\sigma_{K^+N}$ $\approx$ 10 mb. The absorption cross 
for the $K^-$ is much larger, $\sigma_{K^-N}$ $\approx$ 50 mb, and thus 
the $K^-$ dynamics can be expected to be dominated by shadowing effects 
\cite{wang98b}. 
\begin{figure}[h]
\begin{center}
\leavevmode
\epsfxsize = 18cm
\epsffile[0 100 550 430 ]{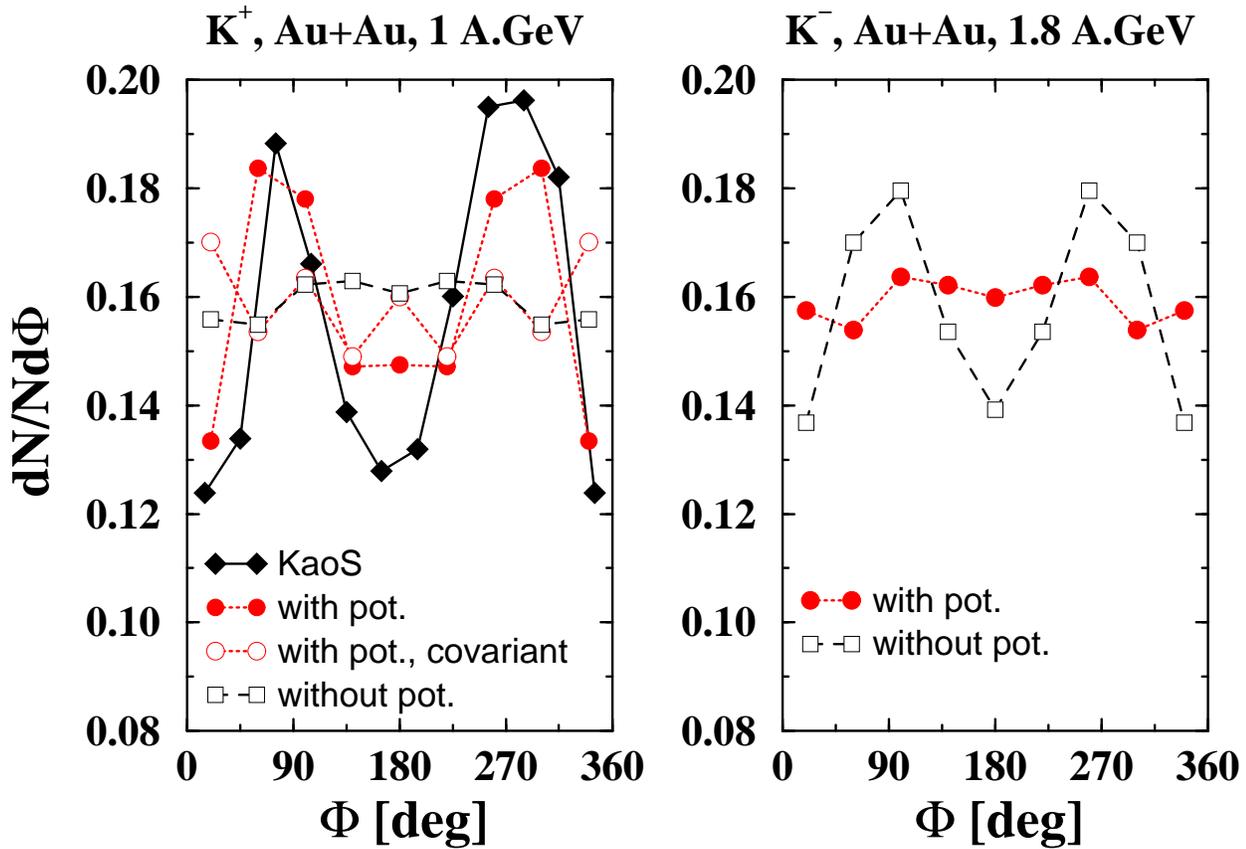}
\end{center}
\caption{{\it Influence of the in-medium potential on 
the kaon squeeze--out. The left panel compares calculations for $K^+$ 
for a semi-central (b=6 fm) Au+Au reaction 
at 1 A.GeV to the KaoS data (diamonds) from \protect\cite{kaos98}. 
The calculations are performed including (full circles) and 
without (squares) the static kaon potential and using full covariant 
kaon dynamics, i.e. the kaon potential with Lorentz 
forces (open circles). The left panel shows calculations for $K^+$ 
for a semi-central (b=8 fm) Au+Au reaction 
at 1.8 A.GeV without (squares) and including the kaon potential 
(full circles).} 
}
\label{squeezefig}
\end{figure}
In Fig. 3 we investigated the squeeze-out for mid-rapidity 
(-0.2 $<$ $(Y/Y_{proj})^{cm}$ $<$ 0.2) $K^+$ mesons in a 
semi-central (b=6 fm) Au+Au reaction at 1 A.GeV incident energy. 
In addition a transverse momentum cut of $P_T > 0.2 $GeV/c has been 
applied in order to compare with the KaoS data 
\cite{senger98}. The QMD calculations are performed for three
different cases: (1) without any in-medium effects, 
(2) including the $K^+$ potential $U (\rho,{\bf k})$, Eq. (\ref{pot}), 
but neglecting the space-like components of the repulsive vector 
potential ${\bf V}$ and (3) with covariant in-medium dynamics, i.e. 
retaining also the space-like vector contribution in Eq. (\ref{lorentz2}). 
The potential is again taken as in Ref. \cite{brown96} (MF2). 
First of all it can be seen that an enhanced out-of-plane emission of 
$K^+$ mesons is mainly a result of the kaon potential. Without any 
medium effects the $K^+$ emission is nearly azimuthally isotropic. 
Very similar effects were also found be the Stony Brook group 
\cite{kaos98,senger98}. 
However, the influence of the space-like components of the repulsive vector 
potential destroys the preferential emission of $K^+$ mesons 
out of the reaction plane and thus also the agreement 
with the data. A similar effect will be also observed for the 
transverse flow of $K^+$ \cite{fuchs98} (see below). 

Concerning $K^-$ mesons the situation is just opposite to the case 
of $K^+$. In the left panel of Fig. 3 the azimuthal 
distributions of $K^-$s emitted at
midrapidity (-0.2 $<$ $(Y/Y_{proj})^{cm}$ $<$ 0.2) in a semi-peripheral 
(b=8 fm) Au+Au reaction at 1.8 A.GeV are shown using a 
$P_T$ cut of $P_T$ $>$ 0.5 GeV. Indeed,  $K^-$s 
are strongly scattered or absorbed in the nuclear medium.
If there is no in-medium potential acting on the $K^-$s, the 
emission at midrapidity exhibits an out-of-plane preference 
much like pions \cite{brill93}. 
Now the in-medium potential (the space-like components 
of the attractive vector part are again neglected) reduces 
dramatically the out-of-plane $K^-$ abundance ($\phi$ = $90^0$ and $270^0$),
and leads thereby to a nearly isotropic emission. Hence, the medium 
effects act opposite on $K^+$ and $K^-$ -- at least in the non-covariant 
description -- and comparing the out-of-plane emission for both 
types of mesons should yield rather conclusive information form 
the experimental side. 
\subsection{Transverse flow}
\begin{figure}[h]
\begin{center}
\leavevmode
\epsfxsize = 18cm
\epsffile[0 100 550 440 ]{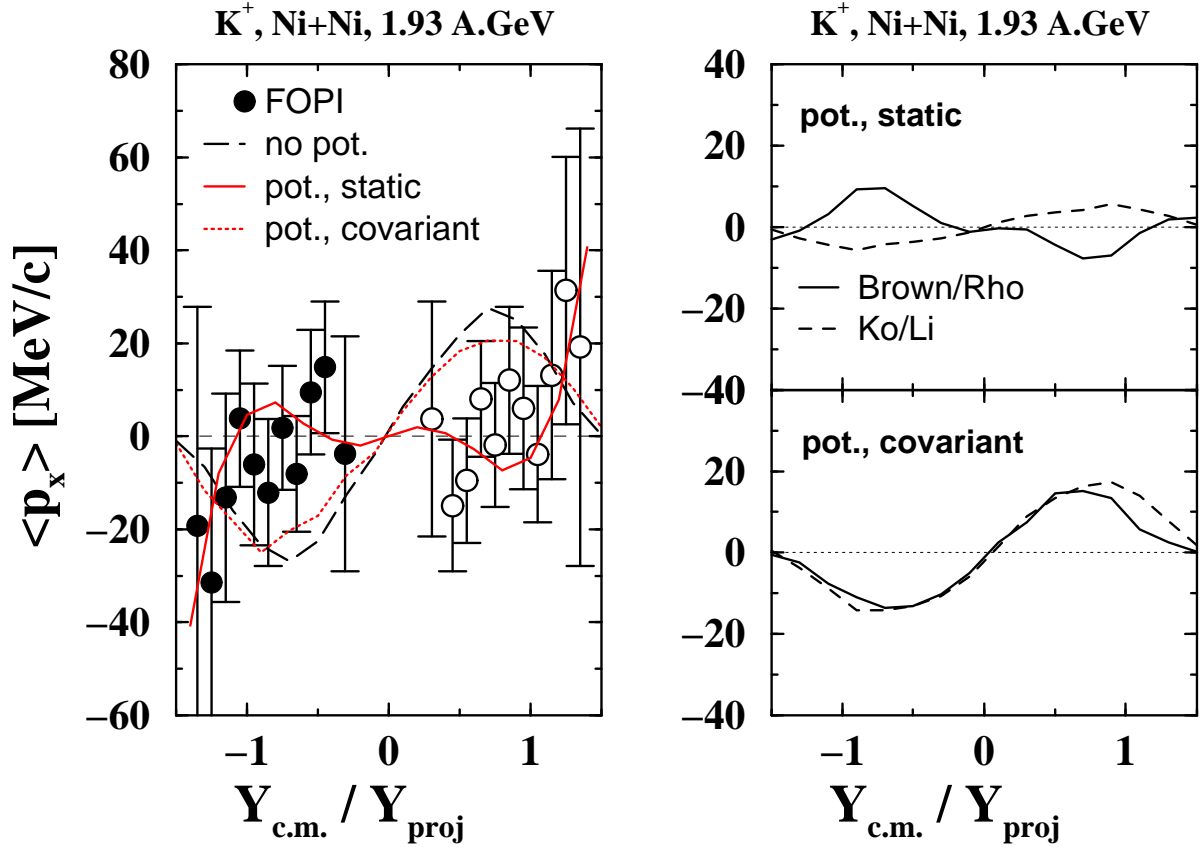}
\end{center}
\caption{{\it Influence of the in-medium potential on 
the $K^+$ transverse flow. The left panel compares calculations for $K^+$ 
for a (b<4 fm) Ni+Ni reaction 
at 1.93 A.GeV to the FOPI data \protect\cite{fopi95}. 
The calculations are performed including (solid) and 
without (dashed) the kaon potential and using full covariant 
kaon dynamics, i.e. the kaon potential with Lorentz 
forces (dotted). The left panel shows the flow for the same reaction at 
b=3 fm for two types of kaon mean fields, again with full covariant 
kaon dynamics (bottom) and using only the potential part (top).}
}
\label{transfig}
\end{figure}
In Fig. 4 the transverse flow of $K^+$ mesons in Ni+Ni collisions at 
1.93 A.GeV is compared to the FOPI data \cite{fopi95}. 
The results are obtained for impact 
parameters $b\leq 4$ fm and with a transverse momentum cut 
$P_{{\rm T}} / m_{{\rm K}} > 0.5$. 
Without medium effects a clear flow signal is observed which reflects 
the transverse flow of the primary sources of the $K^+$ production. 
The dominantly repulsive character of the in-medium potential 
(\ref{pot}) tends to push the kaons away from the spectator matter 
which leads to a zero flow around midrapidity. The situation changes, however, 
dramatically when the full Lorentz structure of the mean field 
is taken into account according to Eqs. (\ref{lorentz},\ref{lorentz2}). 
The influence of the repulsive part of the 
potential, i.e. the time-like component, on the in-plane flow is 
almost completely counterbalanced by the velocity dependent part 
of the interaction. Hence, no net effect of the potential is 
any more visuable. This feature is rather independent 
on the actual strength of the potential as can be seen from the right 
panel of Fig. 4 where the same calculation is performed for 
Ni+Ni at 1.93 A.GeV and an fixed impact parameter b= 3 fm. 
Although the two types of potentials vary considerably in the strength 
of the kaonic vector field which gives rise to significantly 
different results in the case of the quasi-potential treatment 
(top), this effect is completely counterbalanced by the Lorentz-force 
contribution included in the lower figure. 
Although this Lorentz force vanishes in nuclear 
matter at rest, it is clear that this force generally 
contributes in heavy ion collisions. 
Kaons are produced in the early phase of the reaction where the 
relative velocity of projectile and target matter is large. Thus 
the kaons feel a non-vanishing baryon current in the spectator 
region, in particular in non-central collisions.  

The cancelation effects on the flow can be 
understood from Eq. (\ref{lorentz2}). 
The vector field is generally 
proportional to the baryon current 
$ j_\mu = (\rho_B, {\bf u}\rho_B) $ where ${\bf u}$ denotes the 
streaming velocity of the surrounding nucleons. Let us for the 
moment assume that ${\bf u}$ is locally constant, then the 
total contribution of the vector field in Eq. (\ref{lorentz2}) 
can be written as 
$ \mp \frac{3}{8\fps} 
\left( 1 - |{\bf v}| |{\bf u}| \cos\Theta \right) 
\frac{\partial \rho_B }{\partial {\bf q}} 
$. 
Now the angle $\Theta$ between the kaon and the baryon 
streaming velocities determines the influence of the 
Lorentz force. Since in our case and also in the 
calculations of Refs. \cite{li97,ko97a} the $K^+$s initially 
follow the primordial flow of the nucleons we have 
$\cos\Theta \sim 1$ which gives rise to the cancelation. 
However, the value of $\Theta$ and also the magnitude of the 
kaon velocity are also related 
to the rescattering of the  $K^+$ mesons with the nucleons. 
An enhanced rescattering as well as a different primordial 
$K^+$ flow might reduce the cancelation 
effects from the Lorentz force. Hence, the complete description 
of the in-plane $K^+$ flow is still an open question and further 
theoretical studies seem to be necessary \cite{nantes98}.  
\section{Conclusions}
We investigated the influence of chiral mean fields on the dynamics of 
$K$ mesons in heavy ion reactions. To apply such mean fields in the 
framework of relativistic transport theory a covariant 
quasi-particle formulation is used. The vector part of the kaon 
self-energy which is of leading order in the chiral expansion 
and originates form the Weinberg-Tomozawa term is subsummized into 
effective, kinetic four-momenta whereas the scalar part due to 
the next-to-leading order kaon-nucleon-sigma term enters into an 
effective mass. As a consequence of the full relativistic dynamics a 
Lorentz-force term appears which is missing in a description with 
a potential-like force only. In agreement with other works we find that 
the collective motion of kaons is strongly influenced by the mean field 
in dense matter. The comparison with data for the $K^+$ squeeze-out 
and the $K^+$ transverse flow strongly favors the existence of such 
in-medium potentials as predicted form the chiral models. However, 
applying the full relativistic dynamics the description of the data 
is more difficult since we observe strong cancelation effects connected 
to the Lorentz-force contribution. Thus we conclude that the present 
mean field description might be too simple and higher order terms 
from the chiral expansion should be taken into account as well. 
The radial motion of the kaons stemming from the fireball region 
is not affected by such questions, i.e. the Lorentz-forces, 
but gives also rise to a clear signal for the in-medium dynamics. 
Here deviations from the pure thermal spectrum are observed which 
can be attributed to an radial flow of $K^+$ and an "'virtual"', 
inverse radial flow of $K^-$. 

\end{document}